# Testing the transition state theory in stochastic dynamics of a genetic switch

Tomohiro Ushikubo[1], Wataru Inoue[2], Mitsumasa Yoda[1], and Masaki Sasai[1, 2, 3]

[1]Department of Computational Science and Engineering, Graduate School of Engineering, Nagoya University, Nagoya 464-8603, Japan

[2]Department of Complex Systems Science, Graduate School of Information Science, Nagoya University, Nagoya 464-8601, Japan

[3]CREST-Japan Science and Technology Agency, Nagoya 464-8603, Japan

**ABSTRACT** Stochastic dynamics of chemical reactions in a mutually repressing two-gene circuit is numerically simulated. The circuit has a rich variety of different states when the kinetic change of DNA status is slow. The stochastic switching transition between those states are compared with the theoretical estimation of the switching rate derived from the idea similar to the transition state theory. Even though the circuit is kept far from equilibrium, the method gives a consistent explanation of the switching kinetics for a wide range of parameters. The transition state theory-like estimation, however, fails to describe transitions involving the state which has the extremely small numbers of protein molecules.

# 1. Introduction

Recent advancement in biotechnology has facilitated detailed quantitative comparison between experiments and theories on gene expression [1-3]. Such comparison has revealed that the small number nature of molecules in individual cells inevitably brings about the stochastic fluctuation in chemical reactions and hence the number of expressed proteins is strongly fluctuating [4,5].

Complex processes in gene expression can be summarized in a coarse-grained network of reactions among DNA and proteins. See Fig.1 for an example of the reaction scheme. Stochastic fluctuation in such model network has been numerically simulated by using the Gillespie algorithm [6]. It should be useful, however, if we could estimate the rate of stochastic switching between genetic states by using knowledge of the stationary distribution of protein numbers without relying on the extensive stochastic simulations on kinetics. Such analyses of kinetics have been successful in chemical reactions through the transition state theory.

As shown in Fig.1, gene expression is a process which explicitly breaks detailed balance. This is very different from the situation in which the usual transition state theory holds. Even in such a far-from-equilibrium process, however, concepts of the potential energy surface and the transition state are valid when the process is described in one dimension: The drift term of one-dimensional Fokker-Planck equation can be regarded as the derivative of "energy", so that the stochastic switching process is the diffusive motion on the energy surface passing across the "transition state". The question should then arise on whether the similar ideas make sense in multidimensional processes [7,8].

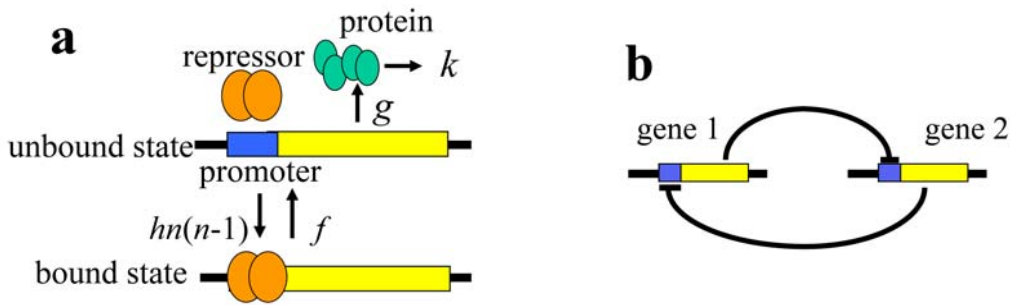

**Figure 1** : Model of a genetic toggle switch.. (a) The model scheme of protein synthesis. Protein molecules are synthesized with the rate $g > 0$ when the promoter dose not bind the repressor and there is no synthesis, otherwise. Proteins are degraded with the rate $kn_i$. The binding rate of the $j$th repressor to the promoter of the $i$th gene is $hn_j(n_j-1)$ with $j = 2$ and 1 for $i = 1$ and 2, respectively and the rate of unbinding is $f$. (b) The genetic toggle switch as a circuit composed of two genes from which repressor proteins are synthesized. Each repressor binds to the promoter of the other gene

Sasai and Wolynes [9] pointed out that interactions in Fig.1 resembles to interactions between electron and surrounding atoms in electron transfer reactions in condensed matter. As in problems in condensed matter, difference in speed of dynamics of constituents should strongly influence the features of the whole system. In prokaryote binding of regulating proteins to DNA takes place in a single cellular compartment, so that the binding state of DNA should change faster than the complex process of protein synthesis. For this reason most of the hitherto developed theories have assumed that the DNA state can quickly reach equilibrium before the other slow events proceed [1,10]. Borrowing the terminology from condensed matter theories, we may be able to call this assumption the "adiabatic" assumption [11-13]. In the strongly adiabatic regime the protein numbers are only the relevant variables and the switching dynamics is represented by a diffusive trajectory in the space of protein numbers. Then, introduction of concepts of the energy surface and the transition-state-theoretical idea might be rather natural. Indeed, the numerical simulation in this regime showed that the transition-state-theoretical estimation of the switching rate is reasonable [14,15].

Recent theoretical analyses showed, however, that the actual cell may not be in the extreme adiabatic limit [12]. In such weakly adiabatic or non-adiabatic case the explicit dynamics of the DNA state plays crucial roles in determining the rate of switching. The problem is essentially multidimensional in this case and it is nontrivial whether the transition state theoretical method is applicable. In this paper we investigate whether the idea of the transition state theory is consistent in such cases.

## 2. Model

We take a mutually repressing two-gene switch as an example, which is often referred to as the genetic toggle switch [16]. As shown in Fig.1 the genetic toggle switch is composed of two operons, each of which consists of the coding region and the promoter region to which the repressor binds. We assume that protein molecules are synthesized with the production rate $g > 0$ when the promoter dose not bind a repressor, and there is no production of proteins when the promoter binds a repressor. Proteins are degraded with the rate $kn_i$ where $n_i$ is the number of protein molecules produced from the $i$th gene. We assume that each repressor works as a dimmer and the dimerization process is fast enough, so that the binding rate of the $j$th repressor to the promoter of the $i$th gene is $hn_j(n_j-1)$ with $j = 2$ and 1 for $i = 1$ and 2, respectively and the rate of unbinding of repressor from the promoter is $f$. We assume for the sake of simplicity that two genes are symmetric and parameters $g$, $k$, $h$, and $f$ do not depend on $i$. Instead of using the bare parameters introduced above, we use normalized ones in simulation; $X = g/(2k)$

represents a typical number of protein molecules, $X^{eq} = h/f$ determines the probability that the DNA state is on, and $\omega = f/k$ is the ratio of the timescale of the DNA state change to the timescale of the protein number change. $\omega$ measures adiabaticity.

Roles of the normalized parameters are clarified when we write down the deterministic version of the rate equation of the model:

$$\frac{1}{k}\frac{dC_i}{dt} = \omega\left((1-C_i) - \frac{N_j^2}{X^{eq}}C_i\right),$$

$$\frac{1}{k}\frac{d}{dt}\left(\frac{N_i}{X}\right) = 2C_i - \frac{N_i}{X}, \qquad (1)$$

where $C_i$ is the probability that the DNA state does not bind the repressor and $N_i$ is the expectation value of $n_i$. In the adiabatic limit of $\omega \gg 1$, the left hand side of the first line of Eq.1 can be neglected, leading to $C_i = X^{eq}/(X^{eq} + N_j^2)$. When we put $C_i = 1/2$ in the second line of Eq.1, we have the stationary solution of $N_i = X$.

The lifetime of proteins should be near to the time span of a cell cycle or less, so that we estimate $k \sim 10^{-2}$-$10^{-3}$s$^{-1}$. Using the data of $\lambda$ phage [17], $f \sim 10^{-1}$-$10^{-3}$s$^{-1}$. Then, $\omega$ should be $\omega \sim 10^{-1}$-$10^{3}$ and $X^{eq} = f/h \sim 10^{2}$-$10^{4}$. We use a typical value of $X^{eq} = 10^{3}$. Typical protein number may be $X \sim 10^{0}$-$10^{3}$. With these parameters the master equation which represents the reaction scheme of Fig.1 is numerically exactly solved by using the Gillespie algorithm [6] and compared with the transition state theoretical estimation.

## 3. Results and discussion

Examples of the time evolution of the protein numbers, $n_1$ and $n_2$, and the corresponding distribution of $n_1-n_2$ are shown in Fig.2. When $X$ is small and $\omega$ is large, fluctuation in the protein number is so large that the trajectory wonders around $n_1=n_2$ and there is no distinct switching behavior. When $X$ becomes larger by keeping $\omega$ large, there appear two relatively stable states with $n_1 > n_2$ and $n_1 < n_2$ and the trajectory can be regarded as a series of switching transitions between them. Hereafter we refer to the protein number realization $(n_1, n_2)$ at the most populated peak of the distribution as "state". When $\omega$ is small, on the other hand, we have a variety of different behaviors and the distribution of $n_1-n_2$ is multi-peaked as shown in Fig.2c-d. The distributions of $n_1$ and $n_2$, $P(n_1,n_2)$, are shown in Fig.3. In Fig.4 the phase diagram is drawn in the $\omega$-$X$ plane by counting the number of peaks in $P(n_1,n_2)$, where peaks of the obtained $P(n_1,n_2)$ with the height larger than 10% of the largest peak are counted but the small ripples in $P(n_1,n_2)$ arising from the numerical sampling problem were neglected. This rich structure of the phase diagram is not expected from the deterministic description of Eq.1 nor from the simulations limited to the adiabatic regime of $\omega \gg 1$ [14,15].

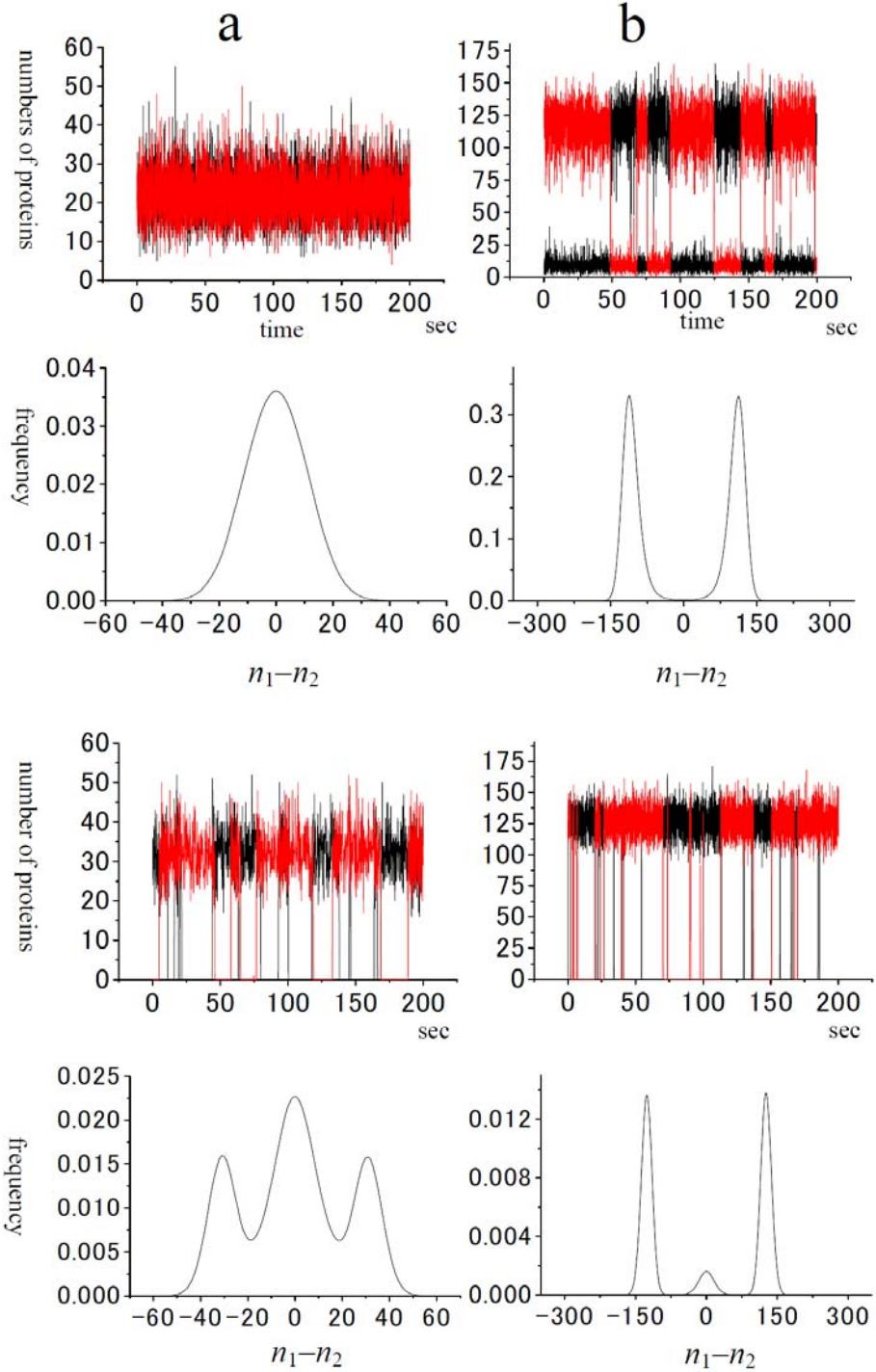

**Figure 2**: Examples of simulated trajectories, $n_1$ and $n_2$ (top), and the distribution $P(n_1, n_2)$ (bottom). In the top figure $n_1$ is red and $n_2$ is black. $X^{eq} = 10^3$ and $k = 10^{-2} s^{-1}$. (a) $\omega = 10^2$, $X = 10^{1.2}$. (b) $\omega = 10^{1.3}$, $X = 10^{1.8}$. (c) $\omega = 10^{-1}$ and $X = 10^{1.2}$. (d) $\omega = 10^{-3}$ and $X = 10^{1.8}$.

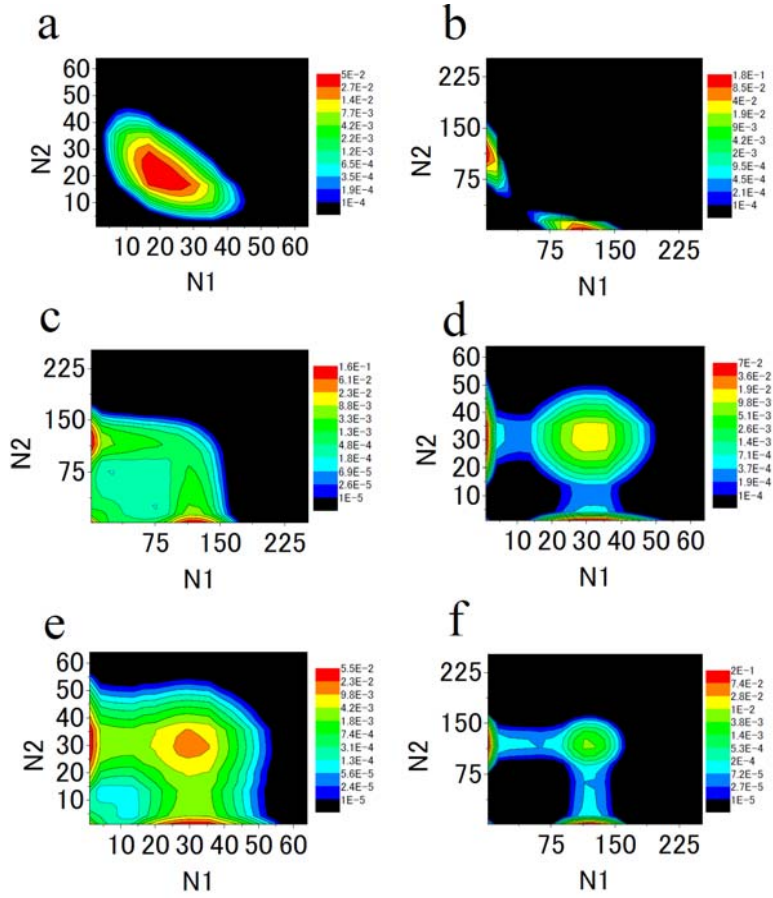

**Figure 3**: Protein number distribution, $P(n_1, n_2)$. Contour maps are colored with the logarithmic scale. $X^{eq} = 10^3$ and $k = 10^{-2} s^{-1}$. (a) $\omega = 10^2$ and $X = 10^{1.2}$, (b) $\omega = 10^2$ and $X = 10^{1.8}$, (c) $\omega = 10^{-1.0}$ and $X = 10^{1.8}$, (d) $\omega = 10^{-2}$ and $X = 10^{1.2}$, (e) $\omega = 10^{-1}$ and $X = 10^{1.2}$, and (f) $\omega = 10^{-3}$ and $X = 10^{1.8}$. (a) to (f) are representative $P(n_1, n_2)$ in regions a to f shown in Fig.4.

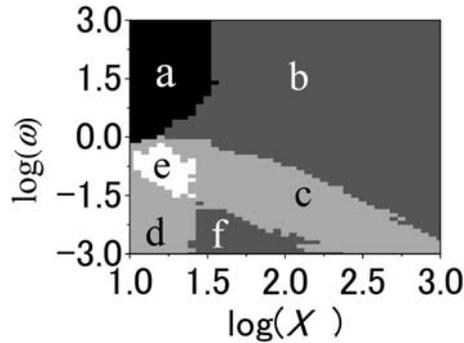

**Figure 4**: Phase diagram of the number of stable states. $X^{eq} = 10^3$ and $k = 10^{-2} s^{-1}$. (a) Monostable, (b) bistable, (c) three stable states at $n_1 > n_2 \approx 0$, $n_1 < n_2 \approx 0$, and $n_1 \approx n_2 \approx 0$, (d) three stable states at $n_1 < n_2 \approx 0$, $n_1 > n_2 \approx 0$, and $n_1 \approx n_2 > 0$, (e) four stable states, and (f) two stable states at $n_1 > n_2 \approx 0$ and $n_1 < n_2 \approx 0$ and an additional weakly stable state at $n_1 \approx n_2 > 0$.

In case that $P(n_1, n_2)$ has a pattern of Fig.5a, for example, we can estimate the rate of transition from the state A at $n_1 > n_2 \approx 0$ to the state B at $n_1 \approx n_2 > 0$ by measuring the time duration needed for the transition, $T_{A\text{-}B}$, along each trajectory. The distribution of $T_{A\text{-}B}$ is shown in Fig.5b. It has a sharp exponential rise within a short time and the slow exponential decrease. The sharp exponential rise implies that there is a minimum time for the transition to take place, which should be a characteristic time of about $1/g$ for $n_2$ to grow from zero. The subsequent slow exponential decay of $\sim \exp(-t/\tau_{A\text{-}B})$ can be interpreted as due to the Poisson process with a characteristic time of $\tau_{A\text{-}B}$. The rate of transition, $k_{A\text{-}B}$, should be estimated by $k_{A\text{-}B} = 1/\tau_{A\text{-}B}$.

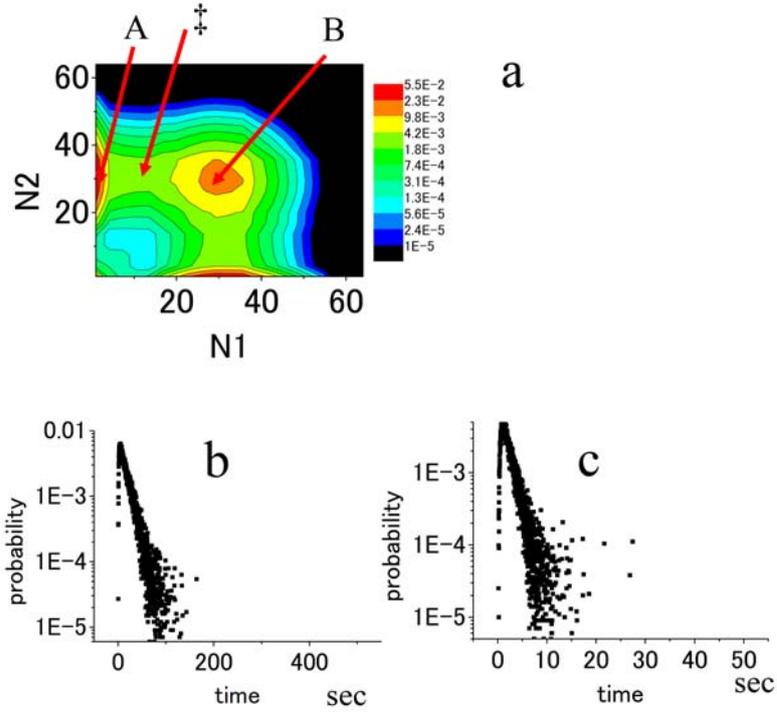

Figure 5: Transition state and distribution of time needed for transition. $X^{eq} = 10^3$ and $k = 10^{-2} \text{s}^{-1}$, $\omega = 10^{-2}$ and $X = 10^{1.2}$. (a) The transition state marked by ‡ is at the saddle of the path connecting the state A and the state B. (b) Distribution of time needed for transition from the state A to the state B, $T_{A\text{-}B}$. (c) Distribution of time needed for transition from the transition state to the state B, $T_{\ddagger\text{-}B}$.

In Ref.15 Warren and ten Wolde measured $k_{A\text{-}B}$ for the transition in the bistable situation of Fig.2b. They showed that $k_{A\text{-}B}$ becomes exponentially small as the protein number becomes large, and suggested that the kinetics is controlled by the factor $\exp(-\beta \Delta F)$, where $\beta \Delta F$ is an extensive quantity proportional to the protein number, which is an analog of free energy in equilibrium systems. A possible interpretation is $\exp(-\beta\Delta F) = P(n_1^\ddagger, n_2^\ddagger)/P(n_1^A, n_2^A)$, where $n_1^A$ and $n_2^A$ are $n_1$ and $n_2$ at the state A and $n_1^\ddagger$ and $n_2^\ddagger$ are $n_1$ and $n_2$ at the separatrix between two basins of state A and B. We may be able to call this separatrix the transition state. Here, the transition state can be identified by defining the coordinate of transition which starts from A to traverse the ridge of $P(n_1, n_2)$ to reach B. The point of the smallest $P(n_1, n_2)$ along the coordinate is the transition state.

In Fig.5c the distribution of $T_{\ddagger\text{-B}}$ is shown, where $T_{\ddagger\text{-B}} = t_{\ddagger\text{-B}} - t_\ddagger$. $t_\ddagger$ is the time when the trajectory passes the transition state and $t_{\ddagger\text{-B}}$ is the time when the trajectory which left the transition state reaches the state B. We can find the characteristic time $\tau_{\ddagger\text{-B}}$ from the exponential relaxation tail of the distribution and estimate the velocity $v_{\ddagger\text{-B}}$ to pass the transition state toward the state B as $v_{\ddagger\text{-B}} = l_{\ddagger\text{-B}}/\tau_{\ddagger\text{-B}}$, where $l_{\ddagger\text{-B}}$ is the length in the ($n_1$, $n_2$) space between the state B and the transition state, $l_{\ddagger\text{-B}} = ((n_1^\ddagger - n_1^B)^2 + (n_2^\ddagger - n_2^B)^2)^{1/2}$. Then, following the transition state theoretical idea, the rate $k_{\text{A-B}}$ should be

$$k_{\text{A-B}} = v_{\ddagger\text{-B}} b^\ddagger \Sigma^\ddagger P(n_1, n_2) / \Sigma^A P(n_1, n_2), \tag{2}$$

where $\Sigma^A$ is summation of $n_1$ and $n_2$ in the basin of the state A. We assumed that the basin is the region around the peak of $P(n_1, n_2)$ within the range that $P(n_1, n_2)$ is larger than $1/e$ of its peak height. $\Sigma^\ddagger$ is summation of $n_1$ and $n_2$ around the saddle point of $P(n_1, n_2)$ along the direction orthogonal to the reaction coordinate. The sum was taken within the range of width $1/e$ around the saddle. $b^\ddagger$ is a transmission factor which is determined by the number of times that a trajectory crosses the transition state in a diffusive way and usually $0 < b^\ddagger < 1$. Since $b^\ddagger$ is determined by the local feature around the transition state, we can expect the rate of the reverse transition should be written with the same $b^\ddagger$ as

$$k_{\text{B-A}} = v_{\ddagger\text{-A}} b^\ddagger \Sigma^\ddagger P(n_1, n_2) / \Sigma^B P(n_1, n_2). \tag{3}$$

Thus, the consistency of the transition state theoretical idea can be checked by calculating $b_A^\ddagger = k_{\text{A-B}} \Sigma^A P(n_1, n_2)/\{v_{\ddagger\text{-B}} \Sigma^\ddagger P(n_1, n_2)\}$, and $b_B^\ddagger = k_{\text{B-A}} \Sigma^B P(n_1, n_2)/\{v_{\ddagger\text{-A}} \Sigma^\ddagger P(n_1, n_2)\}$ from the simulation results. Table 1 summarizes $b_A^\ddagger$ and $b_B^\ddagger$ for several cases. We can find that $0 < b_A^\ddagger \approx b_B^\ddagger < 1$ for transitions between the state of $n_1 > 0$ and $n_2 \approx 0$ and the state of $n_1 > 0$ and $n_2 > 0$. There is some difference between $b_A^\ddagger$ and $b_B^\ddagger$ in these examples but this difference may be attributed to the errors in estimation of $v_{\ddagger\text{-A}}$ or $v_{\ddagger\text{-B}}$, or to the definition of the reaction coordinate which is not fully optimized. Considering such room of improvement, we can regard that the transition state picture is consistent for transitions between the state of $n_1 > 0$ and $n_2 \approx 0$ and the state of $n_1 > 0$ and $n_2 > 0$ in the weakly adiabatic and non-adiabatic regimes. For transitions involving the state of $n_1 \approx n_2 \approx 0$, however, $b_A^\ddagger$ and $b_B^\ddagger$ are largely different as $0 < b_A^\ddagger < 1 << b_B^\ddagger$, implying that the transition state theoretical idea is clearly inconsistent. In such case the discreteness of the protein number becomes so evident when $n_1 \approx n_2 \approx 0$ that the trajectory is far different from the continuous diffusive motion over the "free energy" like surface, which should prevent the transition state theoretical interpretation of the kinetics. The present results showed, however, that the transition state theoretical idea is consistent in the wide parameter region when a suitable correction is made by introducing the transmission coefficient, suggesting that further development of the theory which can treat both adiabatic and non-adiabatic regimes will offer a unified perspective of the gene switching dynamics.

**Table 1** Transition rates and transmission coefficients in transitions between states of the genetic toggle switch in the weakly adiabatic or non-adiabatic regimes.

| $\omega, X$ | State A $n_1^A, n_2^A$ | State B $n_1^B, n_2^B$ | $k_{A-B}/k$ | $k_{B-A}/k$ | $b_A^\ddagger$ | $b_B^\ddagger$ |
|---|---|---|---|---|---|---|
| $10^{-1}, 10^{1.2}$ | 34, 0 | 32, 32 | 0.082 | 0.065 | 0.309 | 0.237 |
| $10^{-2}, 10^{1.2}$ | 34, 0 | 34, 34 | 0.010 | 0.015 | 0.218 | 0.263 |
| $10^{-3}, 10^{2}$ | 200, 0 | 200, 200 | 0.0087 | 0.0008 | 0.498 | 0.779 |
| $10^{-1}, 10^{1.2}$ | 34, 0 | 0, 0 | 0.068 | 0.078 | 0.0741 | 7.27 |
| $10^{-1}, 10^{2}$ | 200, 0 | 0, 0 | 0.0087 | 0.050 | 0.458 | 12.8 |

**Acknowledgement**

This work was supported by grants from the Ministry of Education, Culture, Sports, Science, and Technology, Japan, and from Japan Society for Promotion of Science and by grants for the 21st century COE program for Frontiers of Computational Science.